\documentclass{llncs}

\usepackage{graphicx,color}
\usepackage{amsmath}
\usepackage{amsfonts}
\usepackage{amssymb}
\usepackage{url} 
\usepackage{multicol}
\usepackage{multirow}
\usepackage{setspace} 
\usepackage{aeguill}
\usepackage{xspace}
\usepackage{listings}
\usepackage{ifthen}
\usepackage{verbatim}
\usepackage{float} 
\usepackage{microtype}
\usepackage{mathptmx}
\usepackage{stmaryrd} 
\usepackage{textcomp}
\usepackage{booktabs}
\usepackage{fancyhdr}

\usepackage{subfigure}

\usepackage[mathletters]{ucs}
\usepackage[utf8x]{inputenc}
\usepackage{wasysym}	

\usepackage[bookmarks=false]{hyperref}

\newcommand{\figref}[1]{\hyperref[#1]{Figure~\ref*{#1}}}
\newcommand{\tableref}[1]{\hyperref[#1]{Table~\ref*{#1}}}
\newcommand{\defref}[1]{\hyperref[#1]{Definition~\ref*{#1}}}
\newcommand{\algref}[1]{\hyperref[#1]{Algorithm~\ref*{#1}}}
\newcommand{\lineref}[1]{\hyperref[#1]{Line~\ref*{#1}}}

\newcommand\bench[1]{\textsf{\small #1}}

\newcommand\maxtime{65 seconds }
\newcommand\maxmemory{130\,MB }

\def\set#1{\{ #1 \}}
\newcommand{\setof}[2]{\set{#1_{1},\ldots,#1_{#2}}}
\newcommand{\condset}[2]{\{ #1 \; | \; #2 \}}

\newcommand{\includenormexample}[1]{#1}

\begin{document}

\title{Structural Analysis:\\
Shape Information via Points-To Computation}

\author{Mark Marron}

\institute{IMDEA Software Institute, \email{mark.marron@imdea.org}}

\maketitle

\begin{abstract}
This paper introduces a new hybrid memory analysis, \emph{Structural Analysis},
which combines an expressive shape analysis style abstract domain with efficient
and simple points-to style transfer functions. Using data from empirical
studies on the runtime heap structures and the programmatic idioms used in modern
object-oriented languages we construct a heap analysis with the following
characteristics: (1) it can express a rich set of structural, shape, and sharing
properties which are not provided by a classic points-to analysis and that are
useful for optimization and error detection applications (2) it uses efficient,
weakly-updating, set-based transfer functions which enable the analysis to be
more robust and scalable than a shape analysis and (3) it can be used as the
basis for a scalable interprocedural analysis that produces precise results in
practice.

The analysis has been implemented for .Net bytecode and using this
implementation we evaluate both the runtime cost and the precision of the
results on a number of well known benchmarks and real world programs. Our
experimental evaluations show that the domain defined in this paper is capable
of precisely expressing the majority of the connectivity, shape, and sharing
properties that occur in practice and, despite the use of weak updates, the
static analysis is able to precisely approximate the ideal results. The analysis
is capable of analyzing large real-world programs (over $30$K bytecodes) in less
than \maxtime and using less than \maxmemory of memory. In summary his work presents
a new type of memory analysis that advances the state of the art with respect to
expressive power, precision, and scalability and represents a new area of study
on the relationships between and combination of concepts from shape and
points-to analyses.
\end{abstract}




\section{Introduction}
\label{sec:intro}
Techniques for analyzing the memory structures created and operated on by a
program have generally fallen into two families: Points-To (Alias) Analysis and
Shape Analysis. These approaches lie at far ends of the spectrum of analysis
cost and precision. In particular points-to analyses track very simple
properties, often little more than points-to set information, and the abstract
transfer functions, which simulate the effects of various program statements,
use simple and efficient set operations. At the other end of the spectrum, shape
analyses track a range of rich heap properties and generally utilize
computationally complex transfer functions, involving materialization
operations, case splitting, and strong
updates~\cite{SagivRW96,GhiyaH96,SagivRW99,cav2008,dillig2}. While individually
each of these areas has seen intensive research, the construction of analysis
techniques that combine the expressiveness of shape style heap domains with the
simplicity and efficiency of points-to style transfer functions is an open
problem. A major challenge in constructing such a hybrid analysis is the 
question of: Are strong updates a fundamental
component of a shape style analysis or is it possible to compute precise shape,
sharing, etc. information with an analysis that uses simpler and more efficient
points-to style transfer functions?

Recent empirical work on the structure and behavior of the heap in modern
object-oriented programs has shed light on how heap structures are
constructed~\cite{stationaryfields,mostlyfuncjava}, the configuration of the
pointers and objects in them~\cite{dacapo}, and their invariant structural
properties~\cite{runtimeabs,heapofshapes,dsfinder}. These results affirm several
common assumptions about how object-oriented programs are designed and how the
heap structures in them behave. In
particular~\cite{stationaryfields,mostlyfuncjava,dacapo} demonstrate that
object-oriented programs exhibit extensive \emph{mostly-functional} behaviors:
making extensive use of \emph{final} (or \emph{quiescing}) fields,
\emph{stationary} fields, copy construction, and when fields are updated the new
target is frequently a newer (often freshly allocated) object. The results
in~\cite{runtimeabs,heapofshapes,dsfinder} provide insight into what heuristics
can be used to effectively group sections of the heap based on how they are used
in the program, how prevalent the use of library containers is, and
what sorts structures are built. The results show that in practice
object-oriented programs tend to organize objects on the heap into well defined
groups based on their roles in the program, they avoid the use of linked pointer
structures in favor or library provided containers, and that connectivity and
sharing properties between groups of objects are relatively simple and stable throughout 
the execution of the program.

The information in these empirical studies provide the central design principles
that guide the construction of the heap analysis in this paper. The prevalence
of mostly functional behavior implies that the domain and transfer functions
can, generally, handle writes as weak updates without large precision losses.
However, to precisely handle object initialization and the frequent case of
updating a field to point to a newly (or very recently) allocated object, the
domain should model such objects with extra care. The extensive use of standard
collections and libraries implies that by specializing the analysis to handle
these collections precisely, as in~\cite{dillig,pasteme}, a large portion of the
potentially complex pointer and indexing operations that would otherwise depend
on the analysis performing strong updates can be eliminated.
Finally, given that 
object-oriented programs are not completely functional, there will be cases where the simplified 
abstract transfer functions introduce imprecision. Thus, the abstract heap domain should provide 
strong disjointness and isolation properties between the various parts of the heap. These 
properties serve to both minimize the impact of any imprecision that is introduced and to prevent 
cascading of this imprecision. 

The \emph{Structural Analysis} abstract domain (\autoref{sec:baselssg}) is based 
on the classic storage shape graph approach and is able to express a rich set of commonly 
occurring and generally useful properties including, structure identification, connectivity, 
sharing, and shape. Additionally, due to the implicit disjointness information in the graph 
structure, the resulting abstract heap model possess strong separability and isolation 
characteristics that limit the propagation 
of imprecision. The normal form (\autoref{sec:normalform}) is defined in terms of an efficient
congruence closure computation, $O((N + E) * \mathit{log}(N))$ where $N$ is the number of nodes 
in the shape graph and $E$ is the number of edges. This congruence relation 
is based on the structures identified in the empirical studies and enables the analysis 
to rapidly converge to a fixpoint without either a large loss of information on the domain 
properties of interest or the generation of large amounts of irrelevant detail. The points-to 
style transfer functions (\autoref{sec:transferfuncs}) are based on set-operations and weak 
updates. In practice they precisely model the heap properties of interest and are efficiently 
computable, $O(N + E)$ worst case but in practice are near constant time. In order to quantify 
the performance and precision of this analysis we present an extensive experimental evaluation 
(\autoref{sec:empevalcase}) of several well known benchmarks including programs from SPEC JVM98 
and DaCapo. This evaluation includes both the timing and memory use characteristics of the 
analysis as well as a rigorous evaluation of the precision of the results. 

\paragraph{Practical Contribution.}
The practical contribution of this paper is the construction of a novel static
heap analysis, \emph{Structural Analysis}, that combines a rich shape analysis
style abstract heap model with efficiently computable points-to analysis style
abstract transfer functions. Our experimental evaluations show that the domain
defined in this paper is capable of precisely expressing the majority of the
connectivity, shape, and sharing properties that occur in practice. Despite the
use of weak updates and the absence of case splitting/materialization the static
analysis is able to precisely (with a rate of $80$-$90$\%) approximate the ideal
results. The memory analysis is, in conjunction with the interprocedural
analysis in~\cite{ipcanalysis}, capable of analyzing real world programs of up
to $30$K bytecodes, which are beyond the capabilities of existing shape
analyses, and never requires more than \maxtime or \maxmemory of memory.

\paragraph{Theoretical Contribution.}
The theoretical contribution of the paper is an answer to the question of the
necessity of strong updates vs. the sufficiency of weak updates in computing
shape and sharing information. The results in this paper show that, despite
previous experience suggesting otherwise~\cite{GhiyaH96,ChaseWZ90a}, strong
updates and the associated machinery are \emph{not critical in practice}, and
that weak updates are sufficient for computing large amounts of useful shape and
sharing information in real world object-oriented programs. This conclusion is
reached via experimental evaluation with the heap analysis constructed in this
paper, \emph{Structural Analysis}, and an analysis of other recent empirical
research~\cite{heapofshapes,dsfinder,stationaryfields,mostlyfuncjava,dacapo}.
Thus, this work opens new possibilities for exploring the relationships between
shape and points-to analyses and represents a new approach to building scalable
and precise memory analysis tools.

%

\section{Abstract Heap Domain}
\label{sec:baselssg}

\newcommand{\predicate}[1]{\textsf{{\small{#1}}}}
\newcommand{\booltype}{\mathsf{bool}}

\newcommand{\Abstraction}{\lambda}
\newcommand{\Concretization}{\gamma}

\newcommand{\ConcEnv}{\ensuremath{\mathsf{Env}}}
\newcommand{\ConcStore}{\ensuremath{\sigma}}
\newcommand{\externalline}{\ensuremath{\ell^{\emph{ext}}}}
\newcommand{\conctype}{\ensuremath{\mathsf{Ty}}}
\newcommand{\concfields}{\ensuremath{\mathsf{Fld}}}
\newcommand{\conclabel}{\ensuremath{l}}
\newcommand{\concaddr}[1]{\ensuremath{a_{#1}}}
\newcommand{\objid}{\ensuremath{\iota}}

\newcommand{\ConcVars}{\ensuremath{\mathsf{Vars}}}
\newcommand{\ConcAddress}{\ensuremath{\mathsf{Addresses}}}
\newcommand{\ConcObs}{\ensuremath{\mathsf{Ob}}}
\newcommand{\Types}{\ensuremath{\mathsf{Types}}}
\newcommand{\ConcLabels}{\ensuremath{\mathsf{Labels}}}
\newcommand{\ObjID}{\ensuremath{\mathsf{OID}}}

\newcommand{\nonep}{\mathsf{none}}
\newcommand{\treep}{\mathsf{tree}}
\newcommand{\dagp}{\mathsf{dag}}
\newcommand{\anyp}{\mathsf{any}}

\newcommand{\AbsEnv}{\ensuremath{\widehat{\mathsf{Env}}}}
\newcommand{\AbsStore}{\ensuremath{\widehat{\sigma}}}
\newcommand{\abstype}{\ensuremath{\widehat{\mathsf{Ty}}}}
\newcommand{\absshape}{\ensuremath{\widehat{\mathsf{Sh}}}}
\newcommand{\absfields}{\ensuremath{\widehat{\mathsf{Fld}}}}
\newcommand{\abslabel}{\ensuremath{\widehat{l}}}
\newcommand{\absaddr}[1]{\ensuremath{\widehat{a_{#1}}}}
\newcommand{\nodeid}{\ensuremath{\widehat{\iota}}}

\newcommand{\absinj}{\ensuremath{\widehat{\mathsf{Inj}}}}
\newcommand{\abstrgts}{\ensuremath{\widehat{\mathsf{Trgts}}}}

\newcommand{\AbsAddress}{\ensuremath{\widehat{\mathsf{Addresses}}}}
\newcommand{\AbsNodes}{\ensuremath{\widehat{\mathsf{Ob}}}}
\newcommand{\AbsLabel}{\ensuremath{\widehat{\mathsf{Labels}}}}
\newcommand{\NodeID}{\ensuremath{\mathsf{NID}}}


We begin by formalizing concrete program heaps and the relevant properties that 
will be captured by the abstraction. Later, we define the abstract heap and formally 
relate the abstraction to the concrete heaps using a \emph{concretization} 
($\gamma$) function from the framework of abstract interpretation~\cite{CC77,NielsonHankinProgAnal}. 
These definitions are designed to support the expression of a range of generally 
useful properties (e.g., shape, sharing, reachability) that are common in 
shape analysis~\cite{GhiyaH96,ChaseWZ90a,MeSharing} and that are useful for a 
wide range of client optimization and error detection applications.

\subsection{Concrete Heaps}
The state of a concrete program is modeled in a standard way where there is an
environment, mapping variables to addresses, and a store, mapping addresses to
objects. We refer to an instance of an environment together with a store and a 
set of objects as a \emph{concrete heap}. Given a program that defines a set of concrete types,
$\Types$, and a set of fields (and array indices), $\ConcLabels$, a concrete
heap is a tuple $(\ConcEnv, \ConcStore, \ConcObs)$ where:
\[
\begin{aligned}
\ConcEnv \in& \text{ Environment } = \ConcVars \rightharpoonup \ConcAddress\\
\ConcStore \in& \text{ Store } = \ConcAddress \rightarrow \text{Objects} \cup \set{\mathsf{null}}\\
\ConcObs \in& 2^{\text{Objects}}\\
\text{Objects} =& \ObjID \times \Types \times (\ConcLabels \rightharpoonup \ConcAddress)\\
& \text{where the object identifier set } \ObjID = \mathbb{N}
\end{aligned}
\] 
Each object $o$ in the set $\ConcObs$ is a tuple consisting of a unique identifier for the object, 
the type of the object, and a map from field labels to concrete addresses for the fields
defined in the object. 
We assume that the objects in $\ConcObs$ and the variables in the environment $\ConcEnv$, 
as well as the values stored in them, are well typed according to the store ($\ConcStore$) 
and the sets $\Types$ and $\ConcLabels$. 

In the following definitions we use the notation $\conctype(o)$ to refer to the type of a given 
object. The usual notation $o.\conclabel$ to refers to the value of the field (or array index) 
$l$ in the object. It is also useful to be able to refer to a \emph{non-null pointer} 
as a specific structure in a number of definitions. Therefore we define a \emph{non-null pointer} 
$p$ associated with an object $o$ and a label as $\conclabel$ in a specific concrete heap, 
$(\ConcEnv, \ConcStore, \ConcObs)$, as $p = (o, \conclabel, \ConcStore(o.\conclabel))$ where 
$\ConcStore(o.\conclabel) \neq \mathsf{null}$. We define a helper function $\concfields : \Types \mapsto 2^\ConcLabels$ 
to get the set of all fields that are defined for a given type (or array indices for an array type).

In the context of a specific concrete heap, $(\ConcEnv, \ConcStore, \ConcObs)$, a
\emph{region} of memory is a subset of concrete heap objects $C \subseteq
\ConcObs$. It is useful to define the set $P(C_1,C_2,\ConcStore)$ of all
non-null pointers crossing from region $C_1$ to region $C_2$ as:
\[ 
\begin{aligned}
P&(C_1, C_2,\ConcStore) = \condset{(o_s, \conclabel, \ConcStore(o_s.\conclabel))}{\exists o_s \in C_1,\conclabel \in \concfields(\conctype(o_s))\,.\,\ConcStore(o_s.\conclabel) \in C_2}
\end{aligned} 
\]
\vspace{-8mm}

\noindent
\paragraph{Injectivity.} 
Given two disjoint regions $C_1$ and $C_2$ in the heap, $(\ConcEnv, \ConcStore,
\ConcObs)$, the non-null pointers with the label $\conclabel$ from $C_1$ to
$C_2$ are \emph{injective}, written
$\mathsf{inj}(C_1,C_2,\conclabel,\ConcStore)$, if for all pairs of non-null
pointers $(o_s, \conclabel, o_t)$ and $(o_s', \conclabel, o_t')$ drawn from
$P(C_1,C_2,\ConcStore)$, $o_s \neq o_s' \Rightarrow o_t \neq o_t'$. As a special
case when we have an array object, we say the non-null pointer set
$P(C_1,C_2,\ConcStore)$ is \emph{array injective}, written,
$\mathsf{inj}_{[]}(C_1,C_2,\ConcStore)$, if for all pairs of non-null pointers
$(o_s, i, o_t)$ and $(o_s', j, o_t')$ drawn from $P(C_1,C_2,\ConcStore)$ and $i$,
$j$ valid array indices, $i \neq j \Rightarrow o_t \neq o_t'$.

These definitions capture the general case of an injective relation being defined from a set 
of objects and fields to target objects. They also capture the special, but important 
case of arrays where each index in an array contains a pointer to a distinct object.

\noindent
\paragraph{Shape.} 
We characterize the shape of regions of memory using standard graph theoretic
notions of trees and general graphs treating the objects as
vertices in a graph and the non-null pointers as defining the (labeled) edge
set. We note that in this style of definition the set of graphs that are trees
is a subset of the set of of general graphs. Given a region $C$ in the concrete 
heap $(\ConcEnv, \ConcStore, \ConcObs)$:

\begin{itemize}
\item The predicate $\anyp(C)$ is true for any graph. We use it as the most general shape that 
doesn't satisfy a more restrictive property.

\item The predicate $\treep(C)$ holds if the subgraph $(C, P(C,C,\ConcStore))$ is acyclic and 
does not contain any pointers that create cross edges.

\item The predicate $\nonep(C)$ holds if the edge set in the subgraph is empty, $P(C,C,\ConcStore) = \emptyset$.
\end{itemize}


\subsection{Abstract Heap}
\label{sec:modeldef}
An abstract heap is an instance of a storage shape graph~\cite{ChaseWZ90a}. More precisely, an 
abstract heap graph is a tuple: $(\AbsEnv, \AbsStore, \AbsNodes)$ 
where:
\[
\begin{aligned}
\AbsEnv \in& \text{ Environments } = \ConcVars \rightharpoonup \AbsAddress\\
\AbsStore \in& \text{ Stores } = \AbsAddress \rightarrow \mathsf{Inj} \times 2^{\text{Nodes}}\\
&\quad \text{where the injectivity values } \mathsf{Inj} = \set{\emph{true}, \emph{false}}\\
\AbsNodes \in& \text{ Heaps } = 2^{\text{Nodes}}\\
\text{Nodes} =& \NodeID \times 2^{\Types} \times \mathsf{Sh} \times(\AbsLabel \rightharpoonup \AbsAddress)\\
&\quad \text{where the shape values, } \mathsf{Sh} =\set{\nonep, \treep, \anyp}\\
&\quad \text{and the node identifier set } \NodeID = \mathbb{N}
\end{aligned}
\]
The abstract store ($\AbsStore$) maps from abstract addresses to tuples consisting of the 
injectivity associated with the abstract address and a set of target nodes. Each node $n$ 
in the set $\AbsNodes$ is a tuple consisting of a unique identifier for the node, a set of types, 
a shape tag, and a map from abstract labels to abstract addresses. The use of 
an infinite set of node identity tags, $\NodeID$, allows for an unbounded number of nodes associated with 
a given type/allocation context allowing the local analysis to precisely represent freshly 
allocated objects for as long as they appear to be of special interest in the program (as defined via the  
normal form, \autoref{sec:normalform}, and used in the transfer functions, \autoref{sec:transferfuncs}) 
The abstract labels ($\AbsLabel$) are the field labels 
and the special label $[]$. The special label $[]$ abstracts the indices of all array elements 
(i.e., array smashing). Otherwise an abstract label $\abslabel$ represents the object 
field with the given name.

As with the objects we introduce the notation $\abstype(n)$ to refer to the type set 
associated with a node. The notation $\absshape(n)$ is used to refer to the shape property, 
and the usual $n.\abslabel$ notation to refer to the abstract value associated with the 
label $\abslabel$. Since the abstract store ($\AbsStore$) now maps to tuples of 
\emph{injectivity} and node target information we use the notation $\absinj(\AbsStore(\absaddr{}))$ 
to refer to the \emph{injectivity} and $\abstrgts(\AbsStore(\absaddr{}))$ to refer to the set of 
possible abstract node targets associated with the abstract address. We define the helper 
function $\absfields : 2^{\Types} \rightarrow 2^{\AbsLabel}$ to refer to the set of all abstract 
labels that are defined for the types in a given set (including $[]$ if the set contains 
an array type). 

\subsection{Abstraction Relation}
\label{sec:absrel}
We are now ready to formally relate the abstract heap graph to its concrete counterparts by specifying which 
heaps are in the concretization ($\Concretization$) of an abstract heap:
\[
 \begin{aligned}
(\ConcEnv, &\ConcStore, \ConcObs) \in \Concretization((\AbsEnv, \AbsStore, \AbsNodes)) \Leftrightarrow \exists \text{ an embedding } \mu \text{ where } \mathsf{Typing}(\mu, \ConcObs, \AbsNodes)\\
&\land \mathsf{Injective}(\mu, \ConcEnv, \ConcStore, \ConcObs, \AbsEnv, \AbsStore, \AbsNodes) \land \mathsf{Shape}(\mu, \ConcEnv, \ConcStore, \ConcObs, \AbsEnv, \AbsStore, \AbsNodes)
\end{aligned}
\]
A concrete heap is an instance of an abstract heap, if there exists an embedding function 
$\mu: \ConcObs \rightarrow \AbsNodes$ satisfying the graph embedding, typing, injectivity, and shape 
relations between the structures. The auxiliary predicates are defined as follows. 
\[
\begin{aligned}
&\mathsf{Embed}(\mu, \ConcEnv, \ConcStore, \ConcObs, \AbsEnv, \AbsStore, \AbsNodes) = \\
& \quad\forall v \in \ConcVars\,.\,\mu(\ConcStore(\ConcEnv(v))) \in \abstrgts(\AbsStore(\AbsEnv(v))) \\
& \quad\land \forall o_s \in \ConcObs \text{ and non-null pointers } p = (o_s, \conclabel, o_t)\\ 
& \quad\qquad \exists \abslabel \in \absfields(\abstype(\mu(o_s)))\,.\,\mu(o_t) \in \abstrgts(\AbsStore(\mu(o_s).\abslabel)) \land \conclabel \in \gamma_{L}(\abslabel)\\
\end{aligned}
\]
The embed predicate makes sure that all of the objects and pointers of the concrete heap are present in the 
abstract heap graph, connecting corresponding abstract nodes, and that the store and labels in the abstract graph 
respect the concrete store and labels. The embedding must also preserve any variable mappings. 
\[
\begin{aligned}
\mathsf{Typing}(&\mu, \ConcObs, \AbsNodes) = \forall n \in \AbsNodes, o \in \mu^{-1}(n)\,.\,\conctype(o) \in \abstype(n)
\end{aligned}
\]
The typing relation guarantees that the type $\conctype(o)$ for every concrete object $o$ is in the set 
of types of the abstract node $\abstype(n)$ associated with $o$.
\[
\begin{aligned}
\mathsf{Injective}&(\mu, \ConcEnv, \ConcStore, \ConcObs, \AbsEnv, \AbsStore, \AbsNodes) = \forall n_s, n_t \in \AbsNodes, \abslabel \in \absfields(\abstype(n_s))\,.\,\absinj(\AbsStore(n_s.\abslabel)) \Rightarrow \\
&(\abslabel \neq [] \Rightarrow \mathsf{inj}(\mu^{-1}(n_s),\mu^{-1}(n_t),\conclabel,\ConcStore)) \land (\abslabel = [] \Rightarrow \mathsf{inj_{[]}}(\mu^{-1}(n_s),\mu^{-1}(n_t),\ConcStore))\\
\end{aligned}
\]
The injectivity relation guarantees that every pointer set marked as injective corresponds to injective 
(and array injective as needed) pointers between the concrete source and target regions of the heap. 
\[
\begin{aligned}
\mathsf{Shape}(&\ConcEnv, \ConcStore, \ConcObs, \AbsEnv, \AbsStore, \AbsNodes) = \forall n \in \AbsNodes\\ &\absshape(n) = \treep \Rightarrow \treep(\mu^{-1}(n,\ConcStore))\land \absshape(n) = \nonep \Rightarrow \nonep(\mu^{-1}(n,\ConcStore))
\end{aligned}
\]
The shape relation guarantees that for every node $n$, the concrete subgraph $\mu^{-1}(n,\ConcStore)$ 
abstracted by node $n$ satisfies the corresponding concrete shape predicates.

\subsection{Example Heap}
\label{sec:exampleheap}
\begin{figure}[t]
	\centering
	\subfigure[A Concrete Heap.]{
		\includegraphics[width=.35\textwidth]{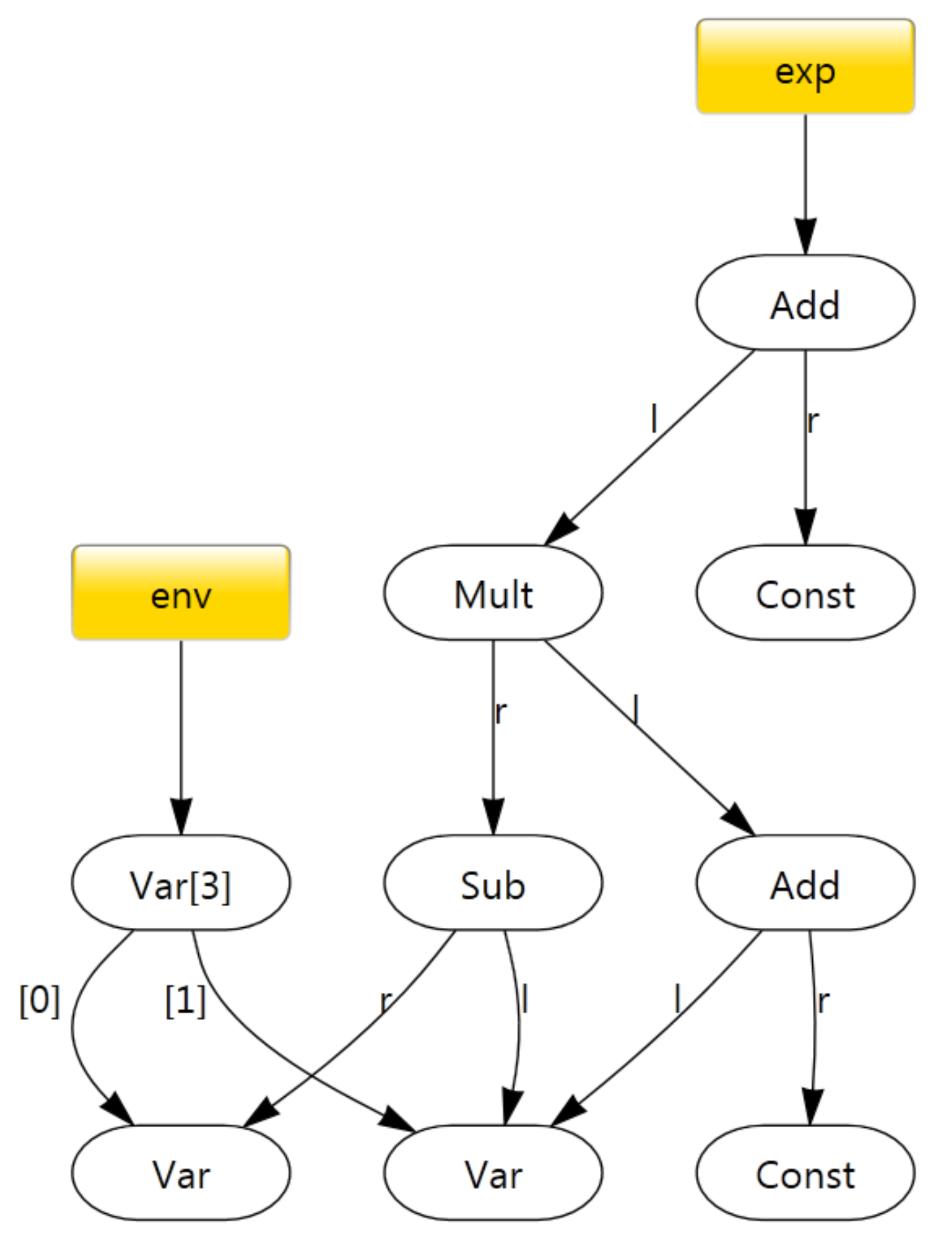}
		\label{fig:concrun}
 	}
 	\subfigure[Corresponding Abstract Heap.]{
 		\includegraphics[width=.5\textwidth]{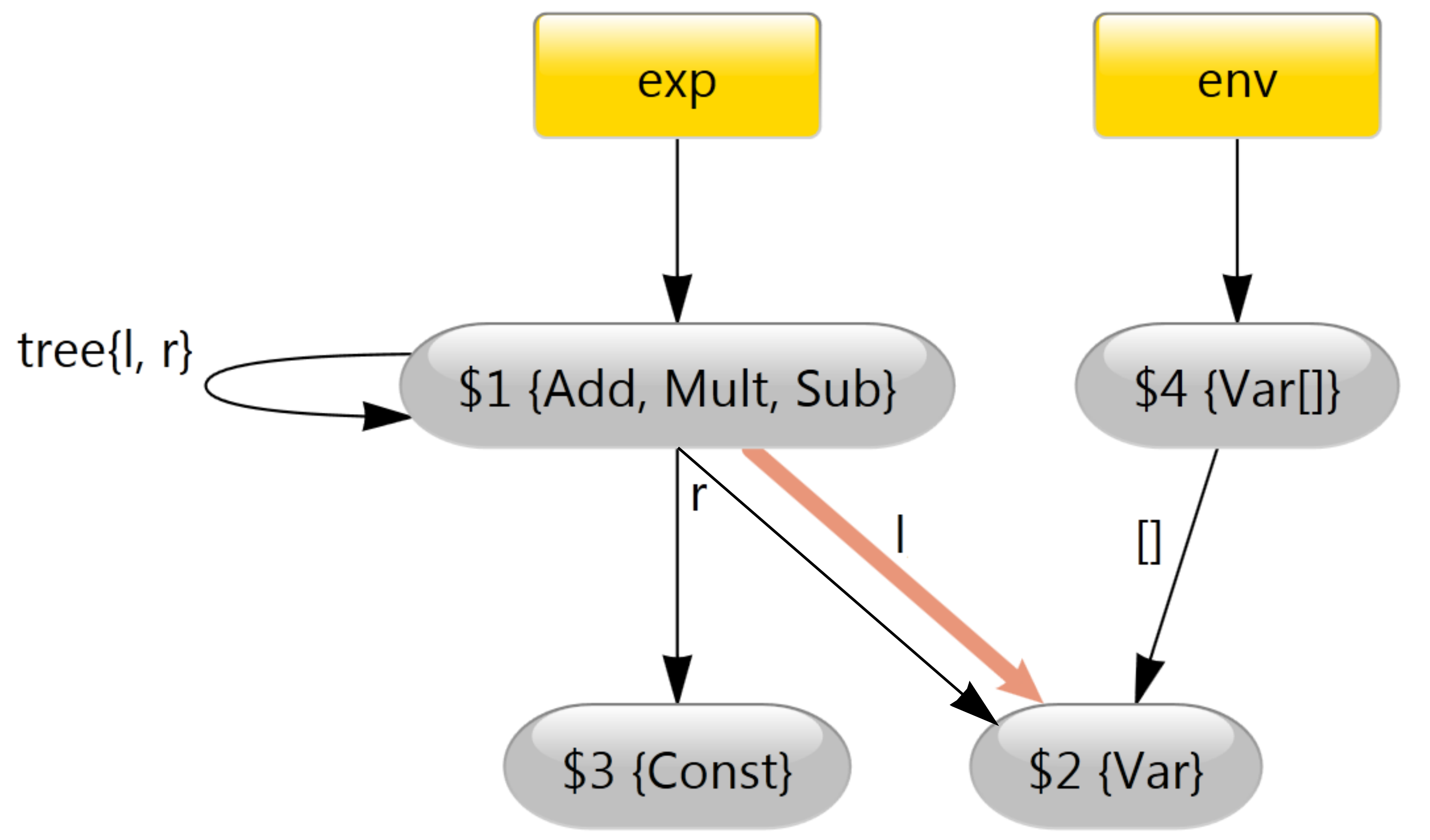}
		\label{fig:absrun}
	 }
\caption{Concrete and Abstract Heap}
\end{figure}

\figref{fig:concrun} shows a snapshot of the concrete heap from a simple program that manipulates 
expression trees. An expression tree consists of binary nodes for \texttt{Add}, 
\texttt{Sub}, and \texttt{Mult} expressions, and leaf nodes for \texttt{Constants} 
and \texttt{Variables}. The local variable \texttt{exp} (rectangular box) points to an 
expression tree consisting of 4 interior binary expression objects, 2 \texttt{Var}, 
and 2 \texttt{Const} objects. The local variable \texttt{env} points to an array representing 
an environment of \texttt{Var} objects that are shared with the expression tree.

%

\figref{fig:absrun} shows the corresponding normal form (see \autoref{sec:normalform}) 
abstract heap for this concrete heap. To ease discussion we label each node in a graph 
with a unique node id (\$id). The abstraction summarizes the concrete objects into three 
regions. The regions are represented by the nodes in the abstract heap graph: 
(1) a node representing all interior recursive objects in the expression tree (\texttt{Add}, \texttt{Mult}, \texttt{Sub}),
(2) a node representing the two \texttt{Var} objects, and
(3) a node representing the two \texttt{Const} objects.
The edges represent possible sets of non-null cross region pointers associated with the given abstract 
labels. Details about the order and branching structure of expression nodes are absent but other 
more general properties are still present. For example, the fact that there is no sharing 
or cycles among the interior expression nodes is apparent in the abstract graph by looking 
at the self-edge representing the pointers between objects in the interior of the expression 
tree. The label \texttt{tree\{l,r\}} on the self-edge expresses that pointers stored in the 
\texttt{l} and \texttt{r} fields of the objects represented by node $1$ form a tree structure.

The abstract graph maintains another useful property of the expression tree, namely that no 
\texttt{Const} object is referenced from multiple expression objects. On the other hand,
several expression objects might point to the same \texttt{Var} object. The abstract graph 
shows this possible non-injectivity using wide orange colored edges (if color is available), 
whereas normal edges indicate injective pointers. Similarly the edge from node $4$ (the \texttt{env} 
array) to the set of \texttt{Var} objects represented by node $2$ is injective, not shaded and wide. 
This implies that there is no aliasing between the pointers stored in the array, i.e. every index 
in the array contains a pointer to a unique object. Additionally, the abstract heap, via a combination 
of reachability, shape, and sharing information, shows there is no aliasing on any distinct pair of 
paths starting from \texttt{exp} and ending with a dereference of the \texttt{r} field. This can 
be deduced from the fact that node 1 is a tree layout, so there is no aliasing internally on either 
the \texttt{l} or \texttt{r} fields, and that both outgoing edges \texttt{r} edges are \emph{injective} 
(narrow and unshaded). Since we know all paths through the tree do not alias (lead to different objects) 
this implies the final dereferences of the \texttt{r} fields, which can only contain injective pointers 
to \texttt{Const} or \texttt{Var} objects, do not alias either.


\section{Normal Form}
\label{sec:normalform}

\newcommand{\nodepartfunc}{\ensuremath{\Pi}}
\newcommand{\nodepart}{\ensuremath{\pi}}

Given the definitions for the abstract heap it is clear that the domain is infinite. This 
allows substantial flexibility when defining the transfer functions and more precise results 
when analyzing straight line blocks of code. However, it is problematic when defining the 
merge/equality operations and can result in the final analysis having an unacceptably large 
computational cost. To prevent this we define an efficiently computable normal form, 
$O((N + E)*\mathit{log}(N))$ where $N$ is the number of nodes in the abstract heap graph 
and $E$ is the number of edges. The normal form ensures that the set of normal form abstract 
heaps for any given program is \emph{finite} and that the abstract heaps in this set can easily 
be merged and compared. 

The normal form leverages the idea that locally (within a basic block or method call) 
invariants can be broken and subtle details are critical to program behavior but before/after these 
local components invariants should be restored. The basis for the normal form, and the selection of 
what are important properties to preserve, comes from studies of the runtime heap structures produced 
in object-oriented programs~\cite{runtimeabs,heapofshapes}. Thus we know that, in general, these definitions 
are well suited to capturing the fundamental structural properties of the heap that are of interest 
while simplifying the structure of abstract heaps and discarding superfluous details. 

\begin{definition}[Normal Form]
We say that the abstract heap is in normal form iff:
\begin{enumerate}
\item All nodes are reachable from a variable or static field.
\item All recursive structures are summarized (\defref{def:samedatastructure}).
\item All equivalent successors are summarized (\defref{def:samelogicalcollection}).
\item All variable/global equivalent targets are summarized (\defref{def:samelogicaltargets}).
\end{enumerate}
That is there are no unreachable nodes and structurally the abstract heap represents 
the congruence closure of the recursive structure, equivalent successor, and equivalent 
target relations.
\label{def:normalform}
\end{definition}

While the normal form definition is fundamentally driven by heuristics derived from 
empirical studies of the heap structures in real programs (and thus one could imagine 
a number of variants) there are three key properties that it possesses: (1) the resulting 
abstract heap graph has a bounded depth, (2) each node has a bounded out degree, and (3) for each 
node the possible targets of the abstract addresses associated with it are unique wrt. the 
label and the types in the target nodes. The first two properties ensure that the number 
of abstract heaps in the normal form set are finite, while the third allows us 
to define efficient merge and compare operations (\autoref{sec:mergecomp}). 

\subsection{Equivalence Partitions}
As each of the properties (\emph{recursive structures}, \emph{ambiguous successors}, and 
\emph{ambiguous targets}) are defined in terms of, congruence between abstract nodes the 
transformation of an abstract heap into the corresponding normal form is fundamentally 
the computation of a congruence closure over the nodes in the abstract heap followed 
by merging the resulting equivalence sets. Thus, we build a map from the abstract 
nodes to equivalence sets (partitions) using a Tarjan union-find structure. Formally 
$\nodepartfunc: \AbsNodes \rightarrow \setof{\nodepart}{k}$ where 
$\nodepart_i \in 2^{\AbsNodes}$ and $\setof{\nodepart}{k}$ are a \emph{partition} of $\AbsNodes$. 
The union-find structure can also be used to maintain the set of all the types associated 
with the nodes in a partition ($\bigcup_{n \in \nodepart} \abstype(n)$). Initially the 
partition is set as a singleton (i.e., $\forall n \in \AbsNodes\,.\,\nodepartfunc(n) = \set{n}$).

The first step in computing the normal form is to identify any nodes that may be parts of 
unbounded depth structures. This is accomplished by examining the type system for the program 
that is under analysis and identifying all the types that are part of the same recursive type 
definitions. This is a commonly used technique~\cite{spelogiccomposite,MYRS05,Deutsch94} 
and ensures that any heap graph produced has a finite depth. We say types $\tau_1$ and $\tau_2$ 
are \emph{recursive} ($\tau_1 \sim \tau_2$) if they are part of the same recursive type definition.

\begin{definition}[Recursive Structure]
Given two partitions $\nodepart_1$ and $\nodepart_2$ we define the \emph{recursive structure} congruence 
relation as:
\[
\begin{aligned}
\nodepart_1& \equiv_r^{\nodepartfunc} \nodepart_2 \Leftrightarrow \exists \tau_1 \in \textstyle{\bigcup_{n_1 \in \nodepart_1}} \abstype(n_1), \tau_2 \in \textstyle{\bigcup_{n_2 \in \nodepart_2}} \abstype(n_2)\,.\,\tau_1 \sim \tau_2\\
&\land \exists n \in \nodepart_1, \abslabel \in \absfields(\abstype(n))\,.\,\abstrgts(\AbsStore(n.\abslabel)) \cap \nodepart_2 \neq \emptyset
\end{aligned}
\] 
\label{def:samedatastructure}
\end{definition}

The other part of the normal form computation is to identify any partitions that have 
\emph{equivalent successors} and variables that have \emph{equivalent targets}. 
Both of these operations depend on the notion of a successor partition which is 
based on the underlying structure of the abstract heap graph: $ \nodepart_1 \text{ a successor of } \nodepart_2 \text{ and } \abslabel \Leftrightarrow \exists n_2 \in \nodepart_2\,.\, \abstrgts(\AbsStore(n_2.\abslabel)) \cap \nodepart_1 \neq \emptyset$.

\begin{definition}[Partition Compatibility]
We define the relation Compatible($\nodepart_1$, $\nodepart_2$) 
as: $\emph{Compatible}(\nodepart_1, \nodepart_2) \Leftrightarrow \bigcup_{n' \in \nodepart_1} \abstype(n') \cap \bigcup_{n' \in \nodepart_2} \abstype(n') \neq \emptyset$.
\label{def:nodesetequiv}
\end{definition}

\begin{definition}[Equivalent Successors]
Given $\nodepart_1$, $\nodepart_2$ which are successors of $\nodepart$ on labels $\abslabel_1$, $\abslabel_2$
we define the relation \emph{equivalent successors} on them as: $\nodepart_1 \equiv_s^{\nodepartfunc} \nodepart_2 \Leftrightarrow \abslabel_1 = \abslabel_2 \land \emph{Compatible}(\nodepart_1, \nodepart_2)$.
\label{def:samelogicalcollection}
\end{definition}

\begin{definition}[Equivalent on Targets]
Given a root $r$ (a variable or a static field) and two partitions $\nodepart_1$, $\nodepart_2$ where $r$ refers to a 
node in $\nodepart_1$ and a node in $\nodepart_2$ we define 
the \emph{equivalent targets} relation as: $\nodepart_1 \equiv_t^{\nodepartfunc} \nodepart_2 \Leftrightarrow \emph{Compatible}(\nodepart_1, \nodepart_2) \land (\text{r is a static field} \lor \nodepart_1,\nodepart_2 \text{ only have local var predecessors})$.
\label{def:samelogicaltargets}
\end{definition}

Using the \emph{recursive structure} relation and the \emph{equivalent successor
(target)} relations we can efficiently compute the congruence closure over an
abstract heap producing the corresponding normal form abstract heap
(\defref{def:samedatastructure}). This computation can be done via a standard
worklist algorithm that merges partitions that contain equivalent nodes and can
be done in $O((N + E)*\mathit{log}(N))$ time where $N$ is the number of abstract
nodes in the initial abstract heap, and $E$ is the number of abstract addresses
in the heap.

\subsection{Computing Summary Nodes}
After partitioning the nodes in the graph with the congruence closure computation we need to merge 
all the nodes in each partition into a summary node. The resulting summary node should safely summarize 
the properties of the all the nodes in the partition. Similarly, we may need to update target and injectivity 
information for the summary nodes in the abstract store. Given a node partition ($\nodepart$) 
that we want to replace with a new summary node ($n_s$) we use the following to compute the abstract 
properties for the summary node and new abstract store $\AbsStore_s$:
\[
\begin{aligned}
&n_s = (\nodeid_{\emph{frsh}}, \sqcup_{\emph{type}}(\nodepart), \sqcup_{\emph{shape}}(\nodepart), \condset{[\abslabel \mapsto \absaddr{\abslabel}]}{\abslabel \in \absfields(\sqcup_{\emph{type}}(\nodepart)), \absaddr{\abslabel} \text{ a fresh address}})\\ 
&\quad \text{where } \nodeid_{\emph{frsh}} \text{ is a fresh node identifier in } \NodeID\\
&\AbsStore_s = \emph{MergeStore}(\AbsStore_s, \abslabel, \nodepart) \text{ for each } \abslabel \in \absfields(\sqcup_{\emph{type}}(\nodepart))\\
&\sqcup_{\emph{type}}(\nodepart) = \bigcup_{n \in \nodepart} \abstype(n)
\end{aligned}
\]
Once this merge is complete we can update the information on the abstract addresses associated 
with each variable in $\AbsEnv$ by replacing any nodes in the target sets with the appropriate 
newly created summary nodes.

\noindent
\paragraph{Shape.} 
The \emph{Shape} information is non-trivial to merge as it depends both on the 
shapes of the individual nodes that are being grouped and also on the connectivity 
properties between them. We first perform a traversal of the subgraph of the partition 
and the (non-self) abstract targets between them. Then based on the discovery of back, 
cross, or tree references (in a graph theoretic sense) and if any of these abstract storage 
location are \emph{not injective} we compute the shape as $\sqcup_{\emph{shape}}(\nodepart)$ 
$=$ $\emph{struct}(\nodepart) \sqcup \bigsqcup_{n \in \nodepart} \absshape(n)$ where :
\vspace{-2mm}
\[
\emph{struct}(\nodepart) = 
\left\{
\begin{array}{cl}
\nonep & \text{ if } \emph{No Internal Edges Exist}\\
\treep & \text{ if } \forall n \in \nodepart, \abslabel \in \absfields(\abstype(n))\,.\, \lnot\absinj(\AbsStore(n.\abslabel)) \land n.\abslabel \emph{ a Tree Edge in } \nodepart\\
&\quad \land (\absshape(n) = \nonep \lor \forall n' \in \abstrgts(\AbsStore(n.\abslabel)) \cap \nodepart\,.\, \absshape(n') = \nonep)\\
\anyp & \text{ otherwise }\\
\end{array} 
\right.
\]
\vspace{-6mm}

\noindent
\paragraph{Injectivity and Abstract Targets.}  
Given a mapping from the partitions to the new summary nodes, 
$\Phi: \emph{Img}(\nodepartfunc) \rightarrow \set{n_{s_1}, \ldots, n_{s_k}}$, 
then for each label, $\abslabel$, and abstract address, $\absaddr{\abslabel}$, that may 
appear in a summary node, $n_s$, we set the values in the abstract store as:
\[
\begin{aligned}
&\emph{MergeStore}(\AbsStore_s, \abslabel, \nodepart) = \AbsStore_s + [\absaddr{\abslabel} \mapsto (\emph{inj}, \emph{trgts})] \text{ where}\\
&\quad \emph{trgts} = \condset{\Phi(\nodepartfunc(n'))}{n' \in \textstyle{\bigcup_{n \in \nodepart}} \abstrgts(\AbsStore(n.\abslabel))}\\
&\quad \emph{inj} = \forall n \in \nodepart\,.\,\absinj(\AbsStore(n.\abslabel)) \land \forall n' \in \nodepart \setminus \set{n}\,.\,\abstrgts(\AbsStore(n_1.\abslabel)) \cap \abstrgts(\AbsStore(n_2.\abslabel)) = \emptyset
\end{aligned}
\]

Injectivity is the logical conjunction of the injectivity of all the source label 
locations, and that the respective targets sets of the nodes that are merged do not 
overlap. In the case where the target sets do overlap, i.e., two distinct nodes 
have abstract labels/addresses that contain the same node, the resulting address 
may not only be associated with injective pointers. Thus, the injectivity value is 
conservatively set to \emph{false} (i.e., \emph{not injective}). The target set is simply 
the remapping of the old nodes in the target sets to the appropriate newly created 
summary nodes.

From the definitions of the summary node computations and the update of the abstract 
store locations the preservation of the safety of the abstraction is straight forward 
to check via case enumeration. It is also clear that each partition is processed 
once in the normal form computation (and similarly the addresses in the abstract store are 
each only visited a constant number of times). Thus, the cost of computing the summaries 
can be done in linear time. Finally, as the congruence closure over given a graph is unique 
the resulting normal form graph, as defined here, is also unique.

\includenormexample{
\subsection{Normal Form on Example Heap}
We can see how this normal form works by using it to transform the concrete heap in 
\figref{fig:concrun} into its normal form abstract representation. This can be done 
by first creating an abstract heap graph that is isomorphic to the concrete heap (i.e., 
create a node for each concrete object and set the appropriate targets in the abstract 
store for each concrete pointer). The resulting isomorphic abstract heap is shown in 
\figref{fig:isoAbs}.

\begin{figure}[h]
\centering
\includegraphics[width=.35\textwidth]{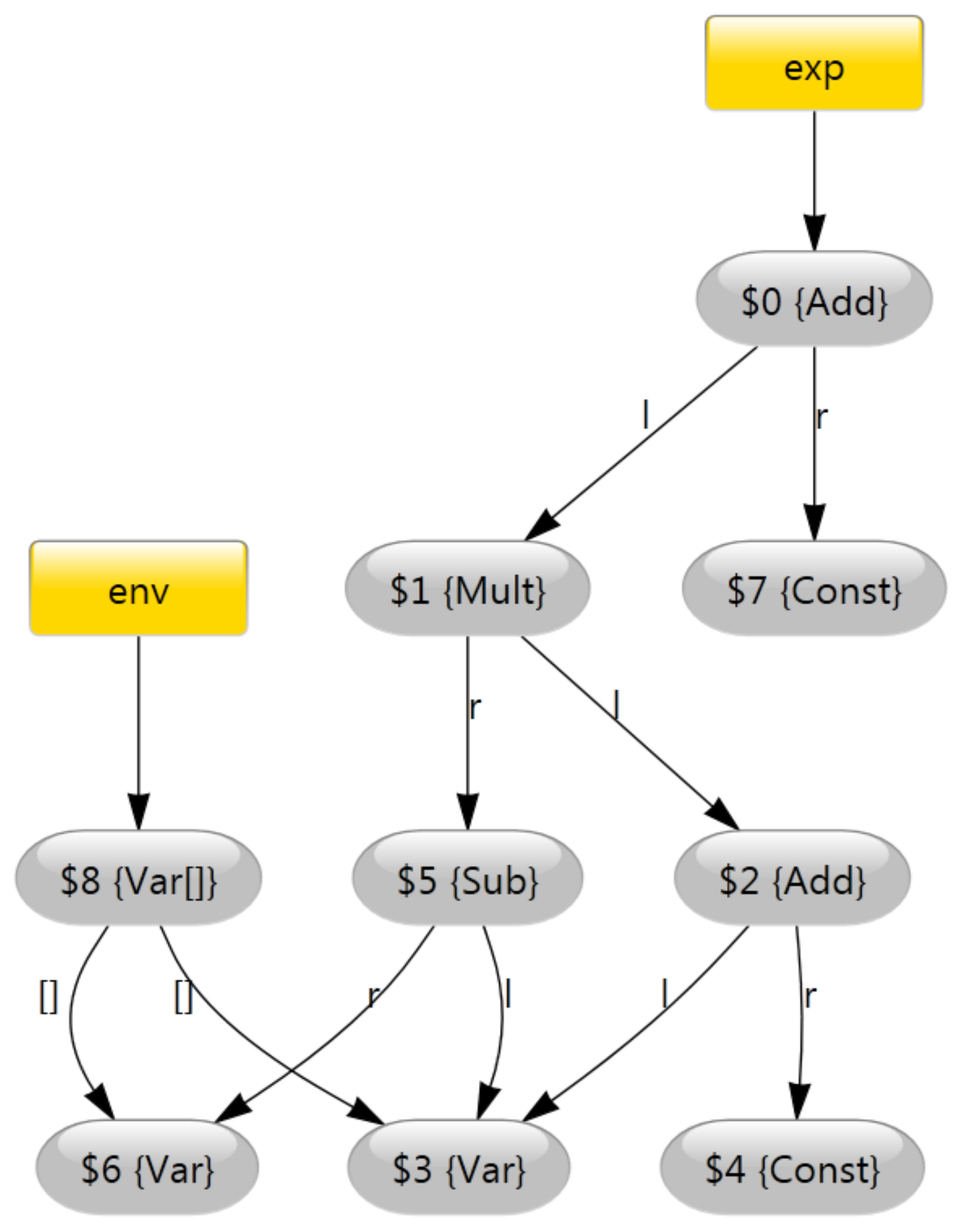}
\caption{Isomorphic Abstract Heap.}
\label{fig:isoAbs}
\end{figure}
}

The normal form partition for the abstract heap in \figref{fig:isoAbs} identifies the 
nodes with the \texttt{Add}, \texttt{Sub}, and \texttt{Mult} types as being in the same 
partition (they are part of the same \emph{recursive structure}). The presence of this 
partition will then cause all of the nodes with \texttt{Const} type (nodes $4$, $7$) to 
be identified as \emph{equivalent successors} of the tree partition. Finally, either due 
to the tree partition or the fact that all the nodes with \texttt{Var} type (nodes $3$, $6$) 
have references to them from node $8$ (the \texttt{Var[]}) will cause all the 
partitions associated with \texttt{Var} types being identified as \emph{equivalent successors}. 
Thus the final partitioning after the congruence closure is:
\begin{displaymath}
\mu^{-1} \left \{
\begin{array}{lr}
\nodepart_1 : \set{n_0, n_1, n_2, n_5} \\ 
\nodepart_2 : \set{n_3, n_6} \\ 
\nodepart_3 : \set{n_4, n_7} \\ 
\nodepart_4 : \set{n_8} \\ 
\end{array}
\right.
\end{displaymath}

Given this set of partitions the computation of the various properties is straight forward. 
The \emph{Shape} for the partitions containing the \texttt{Var}, \texttt{Const} and 
\texttt{Var[]} nodes is trivial to compute as there are no internal references between the 
nodes in these partitions. The \emph{shape} computation for the partition ($\nodepart_1$) 
containing the nodes in the expression structure requires a traversal of the four nodes, 
and as there are no internal cross or back edges the layout for this is $\treep$. 

In computing the new summary abstract store properties for the abstract address associated 
with the expression tree partition ($\nodepart_1$) and the label \texttt{l} there are two nodes 
($n_2$ and $n_5$) that refer to the same node ($n_3$) in partition $\nodepart_2$. Thus this abstract 
storage location is set to not injective (\emph{false}). However, for the label \texttt{r} from 
partition $\nodepart_1$ the target sets are disjoint and thus the injectivity 
in the abstract store is set to \emph{true} (\emph{injective}). Similarly, the store location for 
the label \texttt{[]} out of the partition $\nodepart_4$ representing the targets of the pointers 
stored in the \texttt{env} array is set as \emph{injective}. This results in the normal form abstract 
heap shown in \figref{fig:absrun}.

\section{Domain Operations}
\label{sec:mergecomp}

Given the normal form in \autoref{sec:normalform} we can define an efficiently computable abstract 
equality operation ($\widehat{=}$) and upper approximation ($\widetilde{\sqcup}$) operator on 
the \emph{normal form} abstract heaps. Since the set of normal form abstract heaps is finite (for a 
given program) we do not need a widening operator. Both operations can be performed efficiently, 
$O(N + E)$ for equality and $O((N + E)*\mathit{log}(N))$ for the upper approximation.

\paragraph{Abstract Equality.}
To enable efficient comparison we only define abstract equality on the normal forms of the abstract 
heap states and we ensure it satisfies the property:
\[
(\AbsEnv_1, \AbsStore_1, \AbsNodes_1) \widehat{=} (\AbsEnv_2, \AbsStore_2, \AbsNodes_2) \Rightarrow \gamma((\AbsEnv_1, \AbsStore_1, \AbsNodes_1)) = \gamma((\AbsEnv_2, \AbsStore_2, \AbsNodes_2))
\]
Since the set of normal form abstract graphs we use in the fixpoint computation is finite this is sufficient 
to guarantee termination and safety of the analysis.

Given two abstract heaps $(\AbsEnv_1, \AbsStore_1, \AbsNodes_1)$ and $(\AbsEnv_2, \AbsStore_2, \AbsNodes_2)$ 
we first determine if they are structurally isomorphic (i.e., is there an isomorphism that respects variable 
and field labels), then we check that all abstract node and store properties in $(\AbsEnv_2, \AbsStore_2, \AbsNodes_2)$ 
have the same values in $(\AbsEnv_1, \AbsStore_1, \AbsNodes_1)$ under the isomorphism.

To efficiently compute the needed isomorphism we use a property of the abstract graphs established 
by the normal form definition (\defref{def:normalform}). By this definition we know that 
each node is reachable from a root location (a local variable or a static field), thus if an 
isomorphism exists it can be found by matching from the roots. Further, we know that for 
each abstract address in the store if there is more than one element in the target set then 
each of these targets must have non-overlapping sets of \emph{types} (from the definition 
of \emph{Compatible}, \defref{def:nodesetequiv}). Thus, to compute an isomorphism between 
two graphs we can simply start pairing the local and static roots and then process the abstract 
structure in a breadth first manner, pairing up nodes based on abstract labels and type sets 
of the targets, leading to new pairings. This either results in an isomorphism between the two 
structures, $\phi$, or it reaches a point where no match is possible and fails without backtracking.
If we find an isomorphism $\phi$ then we check the equivalence of the abstract nodes and store as follows:
\[
\begin{aligned}
(\AbsEnv_1&, \AbsStore_1, \AbsNodes_1) =_{\phi} (\AbsEnv_2, \AbsStore_2, \AbsNodes_2) \Leftrightarrow \\
&\forall n\in \AbsNodes_1\,.\,\abstype(n) = \abstype(\phi(n)) \land \absshape(n) = \absshape(\phi(n))\\
&\quad\land \forall \abslabel \in \absfields(\abstype(n))\,.\,\absinj(\AbsStore_1(n)) = \absinj(\AbsStore_2(\phi(n)))
\end{aligned}
\]

\paragraph{Upper Approximation.}
The upper approximation operation takes two abstract heaps and produces a new abstract heap that is 
an over approximation of all the concrete heap states that are represented by the two input abstract 
heaps. In the standard abstract interpretation formulation this is typically the least element that 
is also an over approximation. However, to simplify the computation we do not enforce this property 
(formally we define an \emph{upper approximation} instead of a \emph{join}). Our approach is to leverage 
the existing definitions from the normal form computation in the following steps.

Given two abstract heaps, $(\AbsEnv_1, \AbsStore_1, \AbsNodes_1)$ and $(\AbsEnv_2, \AbsStore_2, \AbsNodes_2)$ 
we can define the their merge by taking the union of the abstract node sets and the abstract stores in the usual way 
and then from this union we can compute the corresponding normal form as described in \autoref{sec:normalform}. 
\[
\begin{aligned}
(\AbsEnv_1&, \AbsStore_1, \AbsNodes_1) \widehat{\sqcup} (\AbsEnv_2, \AbsStore_2, \AbsNodes_2) = \emph{Normalize}(\AbsEnv_m, \AbsStore_m, \AbsNodes_1 \uplus \AbsNodes_2) \text{ where}\\
&\AbsEnv_m = \condset{[v \mapsto \absaddr{v}]}{v \in \emph{Dom}(\AbsEnv_1 \cup \AbsEnv_2), \absaddr{v} \text{ a fresh address}}\\
&\AbsStore_m = \AbsStore_1 \uplus \AbsStore_2 \uplus \condset{[\absaddr{v} \mapsto (\emph{true}, \emph{trgts}_v)]}{[v \mapsto \absaddr{v}] \in \AbsEnv_m}\\
&\quad \emph{trgts}_v = \abstrgts(\AbsStore_1(\AbsEnv_1(v))) \cup \abstrgts(\AbsStore_2(\AbsEnv_2(v)))
\end{aligned}
\]

\section{Abstract Transfer Functions}
\label{sec:transferfuncs}

Given the expressive \emph{Shape Analysis Style} domain defined in \autoref{sec:modeldef} 
the next step is to define a set of transfer functions that simulate the effects of 
various program statements on the abstract heaps. Our goal is to construct these 
definitions in a \emph{Points-To Analysis Style}, using weak updates and simple set operations 
while still precisely modeling the effects of each statement on the 
heap state. In order to focus on the fundamental aspects of the analysis we present the results on 
a simple object-oriented language with the standard set of allocation, load, and store 
operations. However, in practice the approach can be extended in a natural way to handle 
a much richer language. Our implementation for .Net bytecode (\autoref{sec:empevalcase})
handles features such as struct types, references to the stack, pointers to the interior 
of objects, and function pointers.

\tableref{table:transfuncs} shows the transition semantics for the statements that are the most interesting 
from the standpoint of memory analysis (see the companion paper~\cite{ipcanalysis} for a full description 
of how the interprocedural analysis is performed). In order to focus on the central ideas we ignore issues 
with null-pointer dereferences, array out-of-bounds errors, etc. In most cases the abstract transfer 
functions are the natural translations of the concrete semantic operations, and are very similar 
to the set of transfer functions seen in a standard points-to analysis~\cite{objsensitive,typesenspointsto,WilsonL95}. 
However, there are a number of important differences from a standard formulation of points-to analysis transfer 
functions, of particular interest are the \texttt{allocation}, \texttt{load}, and \texttt{store} operations.

\begin{table*}
\begin{tabular}{p{\textwidth}}
\begin{minipage}{\textwidth}
\[
\begin{aligned}
&v = \texttt{alloc type: } (\AbsEnv, \AbsStore, \AbsNodes) \leadsto (\AbsEnv, \AbsStore', \AbsNodes') \text{ where}\\
&\quad n = (\nodeid_{\emph{frsh}}, \texttt{type}, \nonep, \condset{\abslabel \rightarrow \absaddr{\abslabel}}{\abslabel \in \absfields(\set{\texttt{type}}), \absaddr{\abslabel} \text{ fresh address}})\\
&\quad\quad \text{where } \nodeid_{\emph{frsh}} \text{ is a fresh node identifier from } \NodeID\\
&\quad \AbsStore' = \AbsStore + [\AbsEnv(v) \mapsto (\emph{true}, \set{n})] + \condset{[n.\abslabel \mapsto (\emph{true}, \emptyset)]}{\abslabel\in \absfields(\set{\texttt{type}})}\\
&\quad \AbsNodes' = \AbsNodes \uplus \set{n}\\
&\qquad\qquad\qquad\qquad\qquad\qquad\qquad\qquad\qquad\qquad\qquad\qquad\qquad\\
&v = v'.\abslabel \texttt{: } (\AbsEnv, \AbsStore, \AbsNodes) \leadsto (\AbsEnv, \AbsStore', \AbsNodes) \text{ where}\\
&\quad v'_{\emph{trgts}} = \abstrgts(\AbsStore(\AbsEnv(v')))\\
&\quad \AbsStore' = \AbsStore + [\AbsEnv(v) \mapsto (\emph{true}, \bigcup_{n \in v'_{\emph{trgts}}} \abstrgts(\AbsStore(n.\abslabel)))]\\
&\qquad\qquad\qquad\qquad\qquad\qquad\qquad\qquad\qquad\qquad\qquad\qquad\qquad\\
&v.\abslabel = v' \texttt{: } (\AbsEnv, \AbsStore, \AbsNodes) \leadsto (\AbsEnv, \AbsStore', \AbsNodes) \text{ where}\\
&\quad v_{\emph{trgts}} = \abstrgts(\AbsStore(\AbsEnv(v)))\\
&\quad v'_{\emph{trgts}} = \abstrgts(\AbsStore(\AbsEnv(v')))\\
&\quad \forall n \in v_{\emph{trgts}}\,.\,\text{if } n \in v'_{\emph{trgts}} \text{ then }\absshape(n) \leftarrow \anyp\\
&\quad\quad \AbsStore' = \AbsStore + [n.\abslabel \mapsto (\emph{inj}, \abstrgts(\AbsStore(n.\abslabel)) \cup v'_{\emph{trgts}})]\\
&\quad\quad \text{where } \emph{inj} = \absinj(\AbsStore(n.\abslabel)) \land \abstrgts(\AbsStore(n.\abslabel)) \cap v'_{\emph{trgts}} = \emptyset\\
&\qquad\qquad\qquad\qquad\qquad\qquad\qquad\qquad\qquad\qquad\qquad\qquad\qquad\\
\end{aligned}
\]
\end{minipage}
\\
\end{tabular}
\caption{Abstract Semantics}
\label{table:transfuncs}
\end{table*}

\paragraph{Allocate.}
The definition of the allocation operation plays a key role in the functioning of 
the analysis. As opposed to the usual points-to definition which will reuse nodes 
in the abstract heap based on some context token, ranging from simple allocation 
type or line number through sophisticated object-sensitive constructions, our 
definition of the allocation operation always creates a fresh node. In this sense 
the definition closely resembles the constructions used in shape style analyses. 

The creation of a fresh node for each visit to an allocation site is critical to 
allowing the analysis to later model stores into/of this object and the impact on 
injectivity and shape. Any finite naming scheme creates situations where there will 
be spurious reuse of a node, which will cause the loss of injectivity and/or shape 
information (e.g., in the store operation or the normal form summary computation). 
Of course the creation of a new node at each visit to an allocation site creates a 
potential problem with the termination of the analysis as the abstract heap state 
may grow without bound. However, by applying the normal form operation from 
\autoref{sec:normalform} at each control flow join point and at each call site we 
can be sure of the termination of the analysis as the set of graphs that are in 
normal form is finite.

\paragraph{Load.}
The load operation is a simple translation of the concrete semantics where the 
target set that is stored into the variable is the union of the target sets of the 
appropriate fields and objects. However, as a variable location always contains 
a single pointer we can strongly update the target set and set the associated 
\emph{injective} value to \emph{true}.

\paragraph{Store.}
The store operation plays a central role in the analysis as it is where special
care needs to be taken to update the injectivity and shape information. It first
gathers all the possible objects that may be stored into ($v_{\emph{trgts}}$)
and all the possible objects that it may be storing references to
($v'_{\emph{trgts}}$). In the update step we compute new values for the possible
shape, the new target node set, and the new injectivity value. The shape
information is handled by checking if the node we are storing into is in the set
of possible targets. If it is then we may be modifying the shape of the
data structure represented by the node we are updating. It is possible to
perform additional checks to be more precise in updating the shape
information but we simply set the shape to the top value ($\anyp$) in the case
that a self store occurs. If there is no self store then the shape is unchanged.

The update to the abstract store involves taking the union of the old target set and the 
new target set (weakly updating the target set) and computing a new injectivity value. 
There are two cases we need to check to determine the new injectivity value. The first is 
if the old injectivity value was \emph{false}, in which case we leave it as 
\emph{false}. The second is if the new target set and the old target set overlap, 
in which case we cannot guarantee that the address is only associated with injective pointers. 
Again in this case we conservatively set the result as not injective. If neither of these cases 
occur then we mark the abstract address as containing injective pointers (i.e., the 
injective value is \emph{true}).

\section{Implementation and Evaluation}
\label{sec:empevalcase}

To facilitate comparisons with other work we have primarily selected
benchmarks that are direct C\# ports of commonly used Java benchmarks including
programs from Jolden~\cite{joldensuite}, the \bench{db} and \bench{raytracer}
programs from SPEC JVM98~\cite{SPEC98}, and the \bench{luindex} and \bench{lusearch}
programs from the DaCapo suite~\cite{joldensuite}. Additionally we have analyzed
the heap abstraction code, \bench{runabs}, from~\cite{runtimeabs}. In practice we
translate the .Net assemblies to a simplified IR (intermediate representation) which
 allows us remove most .Net specific idioms from the core analysis and 
simplifies later analysis steps. Our test machine is an Intel i7
class processor at $2.66$\,GHz with $2$\,GB of RAM available. We use the
standard $32$ bit .Net JIT and runtime framework provided by Windows $7$. The
domain, operations, and data flow analysis algorithms are all implemented in C\#
and are publicly available.\footnote{Source code and benchmarks available at:
\url{http://jackalope.codeplex.com}}\\

\noindent
{\bf \em Client Applications} \\
\noindent
The analysis in this paper tracks general classes of properties that have shown,
in past work, to be both relevant and useful in a wide range of client
applications~\cite{GhiyaH96,MeSharing,gulwanicav07,cronies,LattnerA05,freeme,unoanalysis}.
However, we have performed additional small scale implementations and case studies 
with the analysis results to ensure that the particulars of the domain defined in this paper 
are useful for these types of optimization and program understanding applications.
These case studies include:
\begin{itemize}
\item The introduction of thread-level parallelization, as in~\cite{MeSharing}, to obtain a $3\times$ speedup for \bench{bh} 
on our quad-core machine. 
\item Data structure reorganization to improve memory usage and automated leak detection, as in~\cite{runtimeabs}, 
to obtain over a $25$\% reduction in live memory in \bench{raytrace}. 
\item The computation of ownership information for object fields, as in~\cite{unoanalysis}, identifying 
ownership properties for $22$\% of the fields and unique memory reference properties for $47$\% of 
the fields in \bench{lusearch}.
\end{itemize}

However, we want to examine the quantitative precision of the analysis in a way 
that is free from biases introduced by the selection of a particular client application. 
Thus, we examine the precision of the static analysis relative to the abstract heaps 
derived from concrete executions of the program. This notion of precision is a 
more general measurement of the possible imprecision due to the use 
of weak-updates and simple points-to style transfer functions than the use of a specific 
client application (which may hide precision losses that \emph{happen} not to matter for the particular client).\\

\noindent
{\bf \em Quantitative Precision} \\
\noindent
We define precision relative to a hypothetical \emph{perfect analysis} which
uses the same abstract domain from \autoref{sec:baselssg} but that is able to
perfectly predict the effects of every program operation. Since we cannot
actually build such an analysis we approximate it by collecting and abstracting
the results of concrete executions. By definition this collection of results
from the concrete execution is an under approximation of the universal
information we want to compute, and in the limit of execution of all possible
inputs is identical. Formally, given a method and a set of concrete heaps
$\setof{h}{k}$ and a set of abstract heaps $\setof{\widehat{h}}{j}$ we can
compute differences between $\bigsqcup_{h \in \setof{h}{k}} \alpha(h)$ and
$\bigsqcup \setof{\widehat{h}}{j}$. This gives an unbiased measure of how close
our results are to the optimal solution, wrt. the abstract domain we are working
with in a way that is independent of peculiarities of a client application or
other analysis technique. 

One possible concern with this approach is that the base abstract domain may be
very coarse, i.e., $\bigsqcup_{h \in \setof{h}{k}} \alpha(h)$ is always $\top$
or another very imprecise value. To account for this we report the average
percentage of properties (shape and injectivity) that the runtime result marks
as precise ($\nonep$ or $\treep$ and \emph{injective}) in the models (the
\emph{Runtime Precise Rate} group in \tableref{tab:precision}). This table shows
that in practice the domain achieves a very high rate of precise identification
of \emph{shape} values, on average over $90$\% or more of nodes are precisely
identified (the \emph{Shape} column), and a similarly high rate of precise
\emph{injectivity} values, on average nearly $90$\% of the edges are identified
as being \emph{injective} (the \emph{Injectivity} column). For reference the
example abstract heap, \figref{fig:absrun}, would have a $100$\% precise \emph{shape} rate and a $75$\%
precise \emph{injectivity} rate. Thus, we can see that in general the base
domain is exceptionally effective in representing the heap properties we are
interested and is an effective baseline for comparison against.

\begin{table}
\centering
\begin{tabular}{| l || c | c | c || c | c |} \hline
\multicolumn{1}{| c || }{} & \multicolumn{3}{c || }{Static Match Rate} & \multicolumn{2}{c | }{Runtime Precise Rate}\\
\hline
Benchmark           & \hspace{1mm} Region \hspace{1mm} & \hspace{1mm} Shape \hspace{1mm} & Injectivity & \hspace{1mm} Shape \hspace{1mm} & Injectivity\\
\hline
\bench{power}        & 100\%  & 100\%  & 100\%     & 100\% & 100\%  \\
\bench{bh}           & 100\%  & 90\%   & 87\%      & 100\% & 100\%  \\
\hline
\bench{db}           & 100\%  & 100\%  & 81\%      & 100\% & 100\%  \\
\bench{raytracer}    & 80\%   & 85\%   & 83\%      & 89\%  & 98\%   \\
\hline
\bench{luindex}      & 95\%   & 95\%   & 82\%      & 100\% & 91\%   \\
\bench{lusearch}     & 93\%   & 90\%   & 84\%      & 96\%  & 89\%   \\
\bench{runabs}       & 97\%   & 98\%   & 87\%      & 94\%  & 90\%   \\
\hline
\end{tabular}
\vspace{3mm}
\caption{Static Match is percentage of each property correctly predicted by the static analysis 
when compared to \emph{perfect analysis}. Runtime Precise is the percentage of properties that 
the \emph{perfect analysis} captures precisely.}
\label{tab:precision}
\end{table}

\tableref{tab:precision} shows the results of comparing the results from our \emph{perfect
analysis} with the results from the static analysis analysis described in this paper. In
this table we report the percentage of properties in the static analysis results
that are the same as reported by the runtime analysis for regions, shapes, and
injectivity values. The region percentage (the \emph{Region} column) is number
of nodes that can be exactly matched between the statically computed and runtime
result structure. Using this matching we then compute the percentage of the
\emph{shape} and \emph{injectivity} properties that are precisely identified by
the static analysis (the \emph{Shape} and \emph{Injectivity} columns). Overall
the results show that the analysis is able to extract a large percentage of the
properties that can be expressed via the selected abstract domain (in general
with a rate of $80$\% to $90$\%). Thus, in general the use of weak-updates and 
points-to style transfer functions result in only small losses in precision 
when analyzing the behavior of the program and the effects of various operations 
on the state of the heap.\\

\noindent
{\bf \em Analysis Performance} \\
\noindent
We next examine the cost of running the analysis in this paper in conjunction with the interprocedural 
analysis described in the companion paper~\cite{ipcanalysis}. For each benchmark we 
list the number of bytecode instructions, the number of classes, and the number of methods that each program 
contains after being translated into the internal IR. These numbers exclude much of the code that would 
normally be part of the runtime system libraries. This is due to the fact that during the translation from 
.Net bytecode to the internal IR code which is never referenced is excluded. Additionally for the builtin 
types/methods that are used the implementations are often replaced by simplified versions or specialized 
domain operations. 

\begin{table}
\centering
\begin{tabular}{| l | l | l | l || r | r |} \hline
\multicolumn{4}{| c || }{Benchmark Statistics} & \multicolumn{2}{c | }{Analysis Cost}\\
\hline
Name	            & Insts \hspace{4mm} & Classes \hspace{2mm} & Methods \hspace{2mm} & \hspace{1mm} Time \hspace{1mm} & \hspace{1mm} Mem \hspace{1mm}\\
\hline
\bench{power}       & 3,298    & 43   & 320   & 0.09s     &  11MB  \\
\bench{bh}          & 3,723    & 45   & 351   & 0.42s     &  14MB  \\
\hline
\bench{db}          & 2,873    & 42   & 315   & 0.21s     &  12MB  \\
\bench{raytracer}   & 9,808    & 65   & 476   & 6.72s     &  32MB  \\
\hline
\bench{luindex}     & 26,852   & 246  & 1747  & 12.1s     &  53MB  \\
\bench{lusearch}    & 33,632   & 272  & 1919  & 64.3s     &  130MB \\
\bench{runabs}      & 27,875   & 253  & 1894  & 10.4s     &  60MB  \\
\hline
\end{tabular}
\vspace{3mm}
\caption{Benchmark statistics and aggregate performance of the analysis.}
\label{tab:performance}
\end{table}

The last two columns of \tableref{tab:performance} show the aggregate performance of the analysis on 
the benchmark set. The timing measurements exclude the time required to startup and read/transform the 
source program into the internal IR. These performance results show that the analysis described in this 
work is quite efficient and capable of analyzing complex programs. 
In the case of \bench{luindex} (a fairly direct translation of the Java version from the DaCapo suite) 
the analysis requires only $12$ seconds while recently reported results on context-sensitive points to 
analyses~\cite{typesenspointsto} reports analysis times ranging between $67$ and $179$ seconds depending 
on the amount and type of object-sensitivity used (and $37$ seconds with an insensitive analysis). But 
more importantly, as memory use frequently is a major scalability wall, are the low memory requirements. 
Despite performing the equivalent of full call graph cloning for large parts of the analysis and being 
partially context sensitive on the remainder, the analysis presented in this paper uses less than \maxmemory 
of memory when analyzing any of the benchmarks.

\section{Related Work}
\label{sec:related}
From the viewpoint of the analysis in this paper work on points-to 
analysis can be seen as falling into two categories, flow-insensitive, and 
flow-sensitive. Flow-insensitive analyses fundamentally 
prioritize speed and scalability over precision and thus are much faster but 
produce much less sophisticated information~\cite{Steensgaard96,antgrasshopper}. In particular 
these approaches can now scale to millions of lines of code with analysis times 
on the order of a few seconds or less~\cite{antgrasshopper}. The second class of 
points-to style analyses are more precise, tracking information in a flow-sensitive 
manner~\cite{Lhotak:2011:PAE,sspfss} and often employing techniques to track information in 
a way that is sensitive to different call sites, either via a context-sensitive or 
object-sensitive approach~\cite{typesenspointsto,objsensitive,JonesMuchnickflow}. 
While these analyses are more precise than flow insensitive points-to analyses they 
do not express general shape or sharing properties. However, due to the way that context 
is tracked they can produce more precise points-to information in some cases than 
the analysis in this paper and it is an open question if object sensitive techniques 
can be used to improve on the results in this paper. Somewhat surprisingly these 
context (or object) sensitive analyses can be slower (and use more memory) than 
the analysis in this paper. 

Work on memory analysis by Latter et. al.~\cite{LattnerAdveDSA,LattnerLA07} is based 
on a modular approach which first builds local shape graphs for each method via a local 
flow-insensitive points to analysis, and then merges (and clones as needed) these local 
graphs via a context-sensitive interprocedural analysis to produce the final result. Due 
to the modular and flow insensitive nature of the analysis it is very efficient, capable 
of analyzing large C++ programs in seconds. The use of a flow-insensitive and a local 
points-to analysis limits the range of properties that can be extracted and the precision 
of the analysis. However, as the focus of this work was scalability (instead of expressivity) 
it provides an interesting contrast in design decisions to the hybrid analysis proposed in 
this paper. Similarly the work in~\cite{HackettR05,dillig2} mixes shape and 
points-to analysis by first partitioning the heap into regions via a flow-insensitive 
points-to analysis followed by performing shape analysis based on these partitions. The work 
of Ghiya and Hendren~\cite{GhiyaH96} uses points-to and basic reachability predicates to 
compute shape information and in Section~4.3 notes the challenges of using weak updates when 
analyzing shape properties. The results in this paper show that the problems 
described in~\cite{GhiyaH96} are not, in practice, fundamental impediments to computing 
precise shape and sharing information in modern object-oriented programs.

There is an extensive body of work on shape 
analysis~\cite{ChaseWZ90a,SagivRW99,GotsmanBC06,cav2008,pasteme,gulwanicav07,dillig2,evanandrival,demskyreachforparallel}, 
and while the work in presented in this paper eschews the use of materialization and 
case splitting in the abstract transfer functions, it borrows heavily from existing work 
in the design of the abstract domain and in the selection of properties it encodes. 
In particular the domain in this paper is based on the basic \emph{storage shape graph} 
construction~\cite{ChaseWZ90a}, which is then augmented with additional information on 
data structure shape~\cite{GhiyaH96} and sharing information (injectivity)~\cite{pasteme}. 
However, as opposed to using a partitioning scheme based on type or allocation site as 
done in~\cite{ChaseWZ90a} (or in most work on points-to analysis) the approach in this paper always creates 
a fresh node in the graph during the \emph{allocation} operation. This node is then grouped 
into other data structures as needed using a normal form operation based on connectivity and 
a set of equivalence relations on the properties of the nodes~\cite{cav2008,MYRS05}. 
The simplicity of the transfer functions in this work, as opposed to the more sophisticated 
shape analysis transfer functions, results in a much faster and more scalable analysis at the 
cost of a small amount of precision. 

We note that existing shape style approaches do not 
currently scale to programs of this size/complexity. While~\cite{biabductive,dillig2} have 
shown promise in scaling to large programs the techniques in~\cite{biabductive} 
do not allow generalized sharing in the heap structures while the techniques 
in~\cite{dillig2} do not handle programs that contain recursive data structures. 
Further, the programs that these approaches have been demonstrated on contain relatively shallow 
and small data structures and the code does not heavily employ recursion or dynamic dispatch. In contrast 
the approach in this paper does not place any constraints on the programs under analysis and 
the benchmarks in this evaluation contain a range of large and complex data structures that are 
connected in a variety of ways. The benchmark programs used in this paper also extensively 
employ dynamic dispatch and contain non-trivial recursive traversals of heap structures.

\section{Conclusion}
\label{sec:conclusion}
This paper introduced \emph{Structural Analysis}, a novel memory analysis technique based on the 
combination of a shape analysis style abstract heap model, a normal form driven by empirical 
studies of heap structures in real-world object-oriented programs, and a set of points-to analysis 
style transfer functions. The resulting hybrid memory analysis is able to precisely identify various 
structures in memory and to track sharing, shape, and reachability relations on them. 
We believe that the combined scalability and precision of the hybrid shape and points-to analysis 
structure presents both immediate benefits and unique opportunities for future research. The development 
of an expressive and scalable heap analysis is a valuable contribution wrt. the wide range of other 
research that depends on information about the program heap. However, we also believe further work 
in the area of hybrid analysis approaches, such as adding object-sensitivity or integrating aspects 
from SMT or separation logic based approaches, will be fruitful areas of investigation. As such we 
believe the analysis presented in this paper represents the introduction of a significant new class 
of heap analysis and represents an important advancement in the state of the art in precise and scalable 
heap analysis techniques.


\bibliographystyle{abbrv}
\bibliography{bibfile}

\end{document}